\documentstyle[12pt]{article}
\newcommand{\be}{\begin{equation}}
\newcommand{\bea}{\begin{eqnarray}}
\newcommand{\eea}{\end{eqnarray}}
\newcommand{\ba}{\begin{array}}
\newcommand{\ea}{\end{array}}
\newcommand{\ee}{\end{equation}}

\expandafter\ifx\csname mathbbm\endcsname\relax

\newcommand{\no}{\nonumber}
\else

\fi
\begin{document}

\begin{titlepage}
\hfill
\vbox{
    \halign{#\hfil         \cr
           ITFA-2001-13 \cr
           hep-th/0104164  \cr
           } 
      }  
\vspace*{20mm}
\begin{center}
{\Large {\bf One-loop Correction of the Tachyon Action in Boundary Superstring 
Field Theory}\\ } 

\vspace*{15mm}
\vspace*{1mm}
{Mohsen Alishahiha }\\

\vspace*{1cm} 

{\it Institute for Theoretical Physics, University of Amsterdam,\\
Valckenierstraat 65, 1018 XE Amsterdam, The Netherlands}\\

\vspace*{1cm}
\end{center}

\begin{abstract}
We compute one-loop correction to the string field theory action of the
tachyon for unstable D-branes in the framework of the boundary superstring 
field theory. We would expect that the one-loop correction comes from the 
partition function of the two-dimensional world-sheet theory on the annulus.
The annulus correction suggests that the genus expansion is, somehow, governed 
by the effective string coupling defined in terms of the tachyon
$\lambda=g_s{\rm exp}(-T^2/4)$.
\end{abstract}
\vskip 6cm

\end{titlepage}

\newpage

\section{Introduction}
The open string tachyon condensation on non-BPS brane systems
 has attracted much interest recently.
One framework of analysis is level truncation of the open string field theory 
(SFT)
which lead to very good numerical agreements with expected values of vacuum energy and lower-dimensional
D-branes tensions \cite{lt,lt1}. Another framework is the boundary SFT (BSFT)
\cite{witten1}-\cite{jap}.  It was argued that while in
the SFT approach an infinite number of  massive fields are 
involved in the condensation process, in the BSFT one can restrict to the tachyon field and study
some aspects of the condensation, such as the tensions of the lower-dimensional D-branes,
 exactly \cite{KMM,KMM2} (see also \cite{lk, jap}).\footnote{ For review of
 tachyon condensation in the open string field theory, see \cite{KO}.} 

The main idea behind BSFT is that `` the space of all two-dimensional 
world-sheet
field theory might be a natural arena for string field theory''. In particular,
the classical configuration space is the space  of two-dimensional world-sheet
theories on the disk which are conformal in the interior of the disk but
have arbitrary boundary interactions. This idea has been first realized for
bosonic string by making use of the Batalin-Vilkovisky 
formalism in \cite{witten1}. This formalism provides a generic expression for the
classical spacetime action in terms of the disk partition function of the
two-dimensional world-sheet theory \cite{witten2, shatashvili1} 
\be
S=(1-\beta^i\frac{\partial}{\partial \lambda^i})\;Z\; ,
\ee
with $\beta^i$ the world-sheet beta function of the boundary coupling 
$\lambda^i$.

It has been recently shown that the Batalin-Vilkovisky 
formalism can also be applied to the
superstring case \cite{M, NP}. In the superstring case, it turns out that
the classical spacetime action is exactly the disk partition function of the
two-dimensional world-sheet theory, $S=Z$. 
We note however that, in the boundary string field theory, it has not been 
shown that the action provides a single cover of the moduli space, which 
was one of the nice things of the cubic open string field theory. Nevertheless,
one hopes that this will somehow be guaranteed by the Batalin-Vilkovisky 
structure. At the same time, one hopes that it will also make sense to
consider this action off-shell, as we are going to compute an off-shall
contribution to the effective action. \footnote{
I would like to thank J. de Boer for bringing my attention to this point and
the point which is discussed in the following paragraph.} 

Regarding the fact that the classical action comes from the disk computation,
one might suspect that the corrections to the action would come from the
surfaces with holes. In this sense, the next correction to the action should
come from the two-dimensional surfaces with zero Euler number which could be
either annulus (orientable) or M\"obius strip (nonorientable). In fact, 
these
corrections can be considered as the quantum effective action that one
would get upon quantizing the classical BSFT action. The quantum effective
action should contain contribution from all higher genus surfaces, so it
can alternatively be computed directly on the world-sheet. 

This is the aim of this note to compute the partition function of the
two-dimensional world-sheet theory on the annulus and, thereby the one-loop 
correction to the tachyon action in the framework of BSFT.

The world-sheet action we are going to study, is 
\be
I=I_0+I_{\rm boundary}\; ,
\label{WORLD}
\ee
with the standard NSR action in the bulk:
\be
I_{0}={1\over 4\pi}\int_{\Sigma} dz^2\left(\partial X^{\mu}{\bar \partial}
 X_{\mu}
+\psi^{\mu}{\bar \partial}\psi_{\mu}+{\bar \psi}^{\mu}\partial \psi_{\mu}
\right)\; .
\label{WSAC}
\ee
The integral is over an annulus $\Sigma$, with inner and outer radii $a$ and
$b$, respectively. 

The boundary is described by superspace coordinates $(\phi,\theta)$, where 
$\theta$ is the boundary Grassman coordinate. Given the boundary superfields  
$\Gamma=\eta+\theta F$ and ${\bf X}=X+\theta \psi $ \cite{WIT}, the boundary action 
$I_{\rm boundary}$ is given by \cite{HKM}:
\be
I_{\rm boundary}=-\int_{\partial \Sigma}\frac{d\phi}{2\pi}d\theta\; 
(\Gamma D\Gamma +T({\bf X})\Gamma)\; ,
\ee
where $D=\partial_{\theta}+\theta\partial_{\phi}$. Performing the integral over 
$\theta$, we get
\be
I_{\rm boundary}=-\int_{\partial \Sigma} 
{d\phi\over 2\pi}\;(F^2+{\dot \eta}\eta+TF+
\psi^{\mu}\eta \partial_{\mu}T) \; .
\ee
The fermions $\eta, \psi $ have anti-periodic boundary condition as is 
appropriate to the NS sector. The boundary auxiliary fields,
$\eta$ and $F$, can be integrated out and the boundary action
reads:
\be
I_{\rm boundary}={1\over 4}\int_{\partial \Sigma} {d\phi\over 2\pi}\left[ T^2+
(\psi^{\mu} \partial_{\mu}T)\partial^{-1}_{\phi} (\psi^{\nu} \partial_{\nu}T)
\right]\; .
\label{INT}
\ee
Here we will consider a case with the following tachyon profile 
\be
T=T_0+u_{\mu} X^{\mu}\; .
\label{TAC}
\ee

The aim is to compute the partition function of the two-dimensional world-sheet
theory (\ref{WSAC}) with the boundary perturbation (\ref{INT}) and 
(\ref{TAC}) and, thereby studying the one-loop correction to the string field 
theory action of the tachyon. We note however that, although it is not clear 
how the boundary string field theory goes beyond the disk approximation,
we will follow the same strategy as the disk for the annulus computation, 
namely, we  assume that the relation between partition function and string 
action, $S[u]=Z[u]$, holds in the annulus approximation. 

We note also that, in general the partition function would be a function of
the annulus modulus, therefore we have to integrate over the annulus modulus.
As a result we have
\be
S[u]=\int_{\rm annulus\;\; modulus} Z[a,b,u] ,
\label{VV}
\ee
where
\be
Z[a,b,u]=\int DX D\psi e^{-I_0-I_{\rm boundary}} .
\label{PF}
\ee

The paper is organized as follows. In section 2, we will explicitly write
down the form of Green's function of the fields $X$ and $\psi$ on the annulus.
In section 3, by making use of the explicit form of the Green's functions, 
we will compute the partition function of the world-sheet theory given by
(\ref{WORLD}) on the annulus and thereby, we shall study the spectrum of open
 strings with different boundary conditions on their end points.
In section 4,
\footnote{I would like to thank J. de Boer for his comment which
led me to add section 4 in order to clarify the misinterpretation of the results in the 
early version of the paper.} we will compute the one-loop 
correction to the string field action of tachyon. In section 5, we shall give
some comment on the results and the problem which this computation
could have. In this paper we shall only consider the case of unstable 
D9-brane. One can easily generalize the result to the other unstable branes. 
An interesting question would be to generalize the results to the cases of
the brane-antibrane system 
and those in the presence of a nonzero B-field as well.

\section{The Green's Function}

In this section we are going to write down the explicit form of 
propagators of $X$ and $\psi$ in the annulus which are needed when we will be
going to compute the partition function (\ref{PF}). 
In general we could have different boundary conditions for each boundary of the
annulus. Therefore the general tachyon profile would be
$T=uX$ at boundary $b$,  and, $T=vX$ at boundary 
$a$. Actually, this means, we are considering an open string stretched between
two D9-branes. In following we set $b=1$.
The Green's function for $X$ is given by  
\cite{TSU} (see also \cite{RVY}):
\bea
G(z,w)&=& -{\rm ln}|z-w|^2+{2\over y}+
{xy\beta\over 2}\;({\rm ln}z\bar z-\frac{2}{y})\;({\rm ln} w\bar w-
\frac{2}{y})
\no\\
&&+\sum_{k=1}^{\infty}{1\over k}C_k\left[
\left(\frac{z}{w}\right)^k+\left(\frac{w}{z}\right)^k+
\left(\frac{{\bar z}}{{\bar w}}\right)^k+
\left(\frac{{\bar z}}{{\bar w}}\right)^k\right]\no\\
&&+\sum_{k=1}^{\infty}\frac{k+y}{k(k-y)}\; C_k
\left[\left(\frac{1}{z\bar{w}}
\right)^k +\left(\frac{1}{\bar z w}\right)^k \right]    \no\\
& &+\sum_{k=1}^{\infty}\frac{k-y}{k(k+y)}\;(C_k+1)
\left[(z{\bar w})^k+({\bar z} w)^k\right]\; ,
\label{GREEN}
\eea
where $y=u^2, \; x=v^2$, and
\be
C_k=\frac{a^{2k}}{\frac{(k+y)(k+ax)}{(k-y)(k-ax)}-a^{2k}}\;, \;\;\;\;\;\;
\beta^{-1}=xy{\rm ln}a-x-{y\over a}\; .
\ee

Setting $z=e^{i\phi}$ and $\omega=e^{i\phi'}$ one can compute the propagator 
of $X$ at the
boundary $b=1$. Collecting all terms together in (\ref{GREEN}), we get
\bea
&&G_{\rm bos}(\phi-\phi';x,y)|_{1}:=\langle X(\phi)X(\phi') \rangle_{(x,y)}
|_{1}\cr
&&\cr
&&=2\sum_{k\in Z}\frac{1}{|k|+y}e^{ik(\phi-\phi')}+2\sum_{k\in Z}
\frac{2|k|}{k^2-y^2}\;C_{|k|}e^{ik(\phi-\phi')}\; .
\eea
Note that in the second summation the $k=0$ case is singular, regularizing this
term we get a term with the form of ${2 x\over y}\beta$.

Doing the same computation for the boundary at $a$, we find:
\be
\langle X(\phi)X(\phi') \rangle_{(x,y)}|_{a}=\langle X(\phi)X(\phi') 
\rangle_{({y\over a},ax)}|_{1}\; .
\ee

In order to find the propagator for the fermion we note the fact that, if the 
fermions $\psi$ were periodic, the contribution of the fermion would precisely
cancel that of the bosons, so the full partition sum would be trivial. 
In fact the inverse derivative of the propagator of the fermion 
 must be very similar to the bosonic one, the only difference being the overall
sign and range of the index $k$ \cite{KMM2}.  Therefore we get 
\bea
&&G'_{\rm fer}(\phi-\phi';x,y)|_1:=\langle \psi(\phi)\partial^{-1}_{\phi'}
\psi(\phi')
\rangle_{(x,y)}|_1\cr
&&\cr
&& =-2\sum_{k\in Z+{1\over 2}}\frac{1}{|k|+y}e^{ik(\phi-\phi')}-2\sum_{k\in Z
+{1\over 2}}
\frac{2|k|}{k^2-y^2}\;C_{|k|}e^{ik(\phi-\phi')}\; ,
\label{FER}
\eea
and for the boundary $a$, we have  
\be
\langle \psi(\phi)\partial^{-1}_{\phi'}\psi(\phi')\rangle_{(x,y)}|_a=
\langle \psi(\phi)\partial^{-1}_{\phi'}\psi(\phi')
\rangle_{({y\over a}, ax)}|_b\; .
\label{FERA}
\ee
Moreover from (\ref{FER}) one can extract the boundary propagator for the 
fermions
\bea
&&G_{\rm fer}(\phi-\phi';x,y)|_1:=\langle \psi(\phi)\psi(\phi')\rangle_{x,y}|_1
\cr
&&\cr
&& =2i\sum_{k\in Z+{1\over 2}}\frac{k}{|k|+y}e^{ik(\phi-\phi')}+2i\sum_{k\in Z
+{1\over 2}}
\frac{2k|k|}{k^2-y^2}\;C_{|k|}\;e^{ik(\phi-\phi')}\; ,
\eea
similarly, from the equation (\ref{FERA}) one can write the boundary propagator
for the fermions at boundary $a$. 
 
By writing the index $k$ in (\ref{FER}) as one half of an odd number, and
rewriting the sum over odd integers as a difference of a sum over all integers
and that over even integers, one finds
\be
G'_{\rm fer}(\phi-\phi'; x, y, a)|_1=G_{\rm bos}(\phi-\phi'; x, y, a)|_1-2 
G_{\rm bos}(\frac{\phi-\phi'}{2}
; 2 x \sqrt{a}, 2y, \sqrt{a})|_1\; ,
\label{REL}
\ee
and the same for the boundary at $a$.


\section{Partition Function}

In this section we shall compute the partition function on the
annulus, by making use of the Green's function given
in the previous section. First we will assume a constant tachyon profile.

The partition function for a constant tachyon profile can be obtained by 
setting $u_{\mu}=0$ in (\ref{TAC}) and performing the 
path integral (\ref{PF}). This leads to
\be
Z[a , 0]=e^{-\frac{1}{8\pi}\int d\phi T_0^2}\int DX D\psi e^{-I_0}
=V_0 e^{-\frac{a+1}{4} T_0^2}\; ,
\ee
where $V_0=\int DX D\psi e^{-I_0}$. Therefore, we get
\be
Z[T_0]=\int_{0}^{1} {da \over a}V_0e^{-\frac{a+1}{4} T_0^2}\; .
\label{PP}
\ee
In order to perform the integral over modulus we need to find $V_0$ as a 
function
of modulus. Indeed, what we need to find is the vacuum amplitude on the annulus.
 For the bosonic open string theory this has been 
computed in \cite{ACNY}. Actually, we need to do the same computation for
the superstring one. The procedure is the same as in \cite{ACNY}, i. e. we 
note that the variation of the one-loop partition function with respect to 
the modulus
is proportional to the expectation value of the energy-momentum tensor 
\cite{ACNY}
\be
\frac{\partial \ln V_0[a]}{\partial a}=
\frac{a}{(1-a^2)\pi}\int d^2z\;({1\over {\bar z}^2} T_{zz}+
{1\over z^2} T_{{\bar z}{\bar z}})\; ,
\ee
where
\be
T_{zz}=-{1\over 2}\langle \partial X \partial X \rangle -{1\over 2}
\langle \psi \partial \psi \rangle
\ee
is the energy-momentum tensor of the two-dimensional world-sheet theory.
A straightforward computation, like what has been done for the bosonic
case in \cite{ACNY}, shows
\be
V_0\sim \prod_{k=1}^{\infty}\frac{1-a^{2k-1}}{1-a^{2k}}\; .
\ee
Plugging this result into (\ref{PP}), one finds
\be
V(T_0)=T_9\int_{0}^{1} {da \over a} \prod_{k=1}^{\infty}
\frac{1-a^{2k-1}}{1-a^{2k}}e^{-\frac{1}{4} T_0^2}\;e^{-\frac{a}{4} T_0^2}\; .
\label{POT}
\ee
Here, in order to emphasize the contribution of each boundary we separated 
the exponential into the two parts. We will use this result for fixing the
integration constant when we will be computing the partition function
for a non-constant tachyon profile. This could also be used for fixing the
the normalization of the coordinates which is important to find the correct
tachyon potential.  

In order to compute the full action, we start with a tachyon profile such that
$T=u X$ at boundary $b=1$ and $T=v X$ at $a$.
So, the boundary action (\ref{INT}) reads:
\bea
I_{\rm boundary}&=&\frac{y}{8\pi}\int d\phi\; 
(X^2+\psi\partial^{-1}_{\phi}\psi) \;\;\;\;\;\;{\rm at}\;\;1 \; ,\cr
&&\cr
I_{\rm boundary}&=&\frac{x}{8\pi}\int d\phi\; 
(X^2+\psi\partial^{-1}_{\phi}\psi) \;\;\;\;\;\;{\rm at}\;\;a\; .
\eea
Plugging the boundary action into (\ref{PF}) and differentiating with respect
to $y$ and $x$, we get
\bea
\frac{\partial {\rm ln}Z[x, y, a]}{\partial y}&=&-{1\over 8\pi}\int 
d\phi\; \langle X^2+\psi\partial^{-1}_{\phi}\psi\rangle \;\;\;\;\;\;{\rm at}
\;\; 1\; ,\cr &&\cr
\frac{\partial {\rm ln}Z[x, y, a]}{\partial x}&=&-{1\over 8\pi}\int 
d\phi\; \langle X^2+\psi\partial^{-1}_{\phi}\psi\rangle \;\;\;\;\;\;{\rm at}
\;\; a \; .
\label{DPF}
\eea
Here the correlator needs to be regularized. In our case the natural
prescription for defining the correlator is \cite{KMM2}
\be
\langle X^2+\psi\partial^{-1}_{\phi}\psi\rangle=\lim_{\epsilon \rightarrow 0}\;
\langle X(\phi)X(\phi+\epsilon)+\psi(\phi)\partial^{-1}_{\phi}\psi(\phi+
\epsilon)\rangle\; .
\label{RE}
\ee
We note that, this regularization has to do separately for both boundaries at 
$a$ and $b=1$. Using the correlators we have obtained in the previous section,
(\ref{RE}) can be written as
\bea
\langle X^2+\psi\partial^{-1}_{\phi}\psi\rangle|_1&=&
\lim_{\epsilon \rightarrow 0}\;
\left[G_B(\epsilon; x, y)+G'_F(\epsilon; x, y)\right]_1\cr
&=& 2\lim_{\epsilon \rightarrow 0}
\;\left[G_B(\epsilon; x, y, a)-G_B(\frac{\epsilon}{2}; 2 x \sqrt{a}, 2y, 
\sqrt{a})\right]_1\; ,\cr
&&\cr
\langle X^2+\psi\partial^{-1}_{\phi}\psi\rangle|_a&=&
\lim_{\epsilon \rightarrow 0}\;
\left[G_B(\epsilon; x, y )+G'_F(\epsilon; x, y )\right]_a\cr
&=& 2\lim_{\epsilon \rightarrow 0}
\;\left[G_B(\epsilon; x, y, a)-G_B(\frac{\epsilon}{2};
2 x \sqrt{a}, 2y, \sqrt{a})\right]_a\; .
\eea  

To evaluate this limit it is convenient to use the following expression
for the bosonic propagator
\bea
G_B(\epsilon; x, y, a )|_1&=&-2{\rm ln}(1-e^{i\epsilon})-2{\rm ln}(1-e^{-i\epsilon})
+{2\over y}-2y\sum_{k=1}^{\infty}\frac{1}{k(k+y)}(e^{ik\epsilon}+e^{-ik\epsilon})
\cr
&&\cr
&+&\frac{2x}{y}\beta+2\sum_{k=1}^{\infty}\frac{2k}{k^2-y^2}\;C_k\;
(e^{ik\epsilon}+e^{-ik\epsilon})\; .
\eea

By making use of this form, we find 
\be
{1\over 2}\langle X^2+\psi\partial^{-1}_{\phi}\psi\rangle|_1=
-4{\rm ln}2+f(y)+g( x, y, a)-\left(f(2 y)+g( 2 x \sqrt{a}, 2y, 
\sqrt{a})\right)\; ,
\ee
where
\bea
f(y)&=&{2\over y}-4\sum_{k=1}^{\infty}\frac{y}{k(k+y)},\cr &&\cr
g( x, y, a )&=&\frac{2x}{y}\beta+4\sum_{k=1}^{\infty}\frac{2k}{k^2-y^2}\;
C_k\; .
\eea
The same expression for the boundary $a$ can be obtained by replacing
$y\rightarrow ax$ and $x\rightarrow y/a$.

Now we have all information to compute the partition function. Using this 
information, the derivative of the partition function, (\ref{DPF}),
for the case of $y\neq ax$ reads
\bea
\frac{\partial {\rm ln}Z[x, y, a]}{\partial y}|_1&=&
{\partial\over\partial y}\ln
\left[|x y\ln a-{y\over a}-x |^{-{1\over 2}} \right.\cr
&&\left. \right.\cr
 &\times& \left. 
\prod_{k=1}^{\infty}\frac{(k+2 y)(k+2 a x)-(k-2 y)(k-2 a x)
a^k}
{\left[(k+y)(k+a x)-(k-y)(k-a x)a^{2k}\right]^2}\right]\; ,\cr&&\cr
\frac{\partial {\rm ln}Z[x, y, a ]}{\partial x}|_a&=&
{\partial\over\partial x}\ln
\left[|x y\ln a-{y\over a}-x |^{-{1\over 2}} \right.\cr
&&\left. \right.\cr
 &\times& \left. 
\prod_{k=1}^{\infty}\frac{(k+2 y)(k+2 a x)-(k-2 y)(k-2 a x)
a^k}
{\left[(k+y)(k+a x)-(k-y)(k-a x)a^{2k}\right]^2}\right]\; .
\eea
Note that the derivatives of the $\ln Z$ with respect to $y$ and $x$ have the
same form and moreover, the expression is symmetric under exchanging of 
$y\leftrightarrow ax$. 
This is the reason why in our computation we excluded the case
of $y=ax$. We will come back to this point later. One can now integrate and 
find the partition function. Using the $\zeta$-function regularization, we get
\be
Z[x, y, a]=Z_0[a] \frac{F(y)F(ax)}{\sqrt{|x y\ln a-{y\over a}-x |}}
\;\frac{\prod_{k=1}^{\infty} (1+C_k)}{\prod_{r={1\over 2}}^{\infty} (1+C_r)}\, ,
\label{PART1}
\ee
where 
\be
F(z)=4^z\; \frac{z\Gamma^2(z)}{2\Gamma(2z)}\, .
\ee  
Moreover, $Z_0[a]$ is the integration constant which in general, as it is written 
explicitly, could be a function of the annulus modulus. It can be fixed by
comparison with the partition function of a constant tachyon 
profile (\ref{POT}). In fact, it can be seen that $Z_{0}[a]=Z_0 a^{-1/2}$ with
a numerical factor $Z_0$. Thus the equation (\ref{PART1}) reads
\be
Z[x, y, a]=Z_0\frac{F(y)F(ax)}{\sqrt{|ax y\ln a-y-ax |}}
\;\frac{\prod_{k=1}^{\infty} (1+C_k)}{\prod_{r={1\over 2}}^{\infty} (1+C_r)}\, ,
\label{PART}
\ee 
We would like to note that, the partition function (\ref{PART}) is a 
monotonically decreasing function of $y$ and $x$ as is needed in order to be
identified with tachyon action. This can be seen if we compare the expression
(\ref{PART}) with that of the disk computation \cite{KMM2}
\be
Z_{\rm disk}[y]=Z_{0} y^{-{1\over 2}} F(y)\; ,
\ee 

Of course it is not the end of story. One  has to integrate over the modulus
of the annulus  
\be
Z[x , y]=Z_0\int_0^1 {da\over a}\;\frac{F(y)F(ax)}{\sqrt{|ax y\ln a-y-ax |}}
\;\frac{\prod_{k=1}^{\infty} (1+C_k)}{\prod_{r={1\over 2}}^{\infty} (1+C_r)}\, .
\label{TT}
\ee

We would like to note that, having an annulus would naturally lead to a 
system with two parallel D-branes corresponding to two boundaries of the
annulus. 
Depending on which boundary condition we would like to impose at the
boundaries of the annulus, we could get different
boundary conditions at the end points of the open strings.
The different
boundary conditions can be emerged by different possibilities which could be
chosen for $x$ and $y$. In particular, setting $y=x$ corresponds to those
open strings with the same boundary condition
at two end points. In this case, the partition function  has the 
following expansion around $y=0$
\footnote{we note that, since the modulus $a$ takes value 
between zero and one, the $\ln a$ term is always negative. As the result 
$(1+a)-ax\ln a$ can not be zero. Therefore the partition function is not
singular. I would like to thank T. Suyama for bringing my attention to 
this point.}
\be
Z[x]=Z_0\int_0^1 {da\over a} 
\prod_{k=1}^{\infty}\frac{1-a^{2 k-1}}{1-a^{2 k}}
\;{1\over \sqrt{ (1+a) x}}\left[ 1+\left(2 (1+a)\ln 2 
+{1\over 2}\;\frac{a \ln a}{1+a}\right) x +\cdots\right]
\ee
which could be used to obtain the string field theory action up to 
two-derivative
\be
S=T_9\int d^{10}X \int_{0}^{1} {da\over a} 
\prod_{k=1}^{\infty} \frac{1-a^{2 k-1}}{1-a^{2 k}} 
e^{-{1+a \over 4}T^2}\left[1+\left(2 (1+a)\ln 2+
\frac{a \ln a}{2(1+a)}\right)
\partial^{\mu}T\partial_{\mu} T\right].
\label{HAD}
\ee
We note, however, that this is not the one-loop correction to the tachyon 
action we are looking for. In fact, it would be wrong if we read the 
one-loop
correction to the tachyon potential using this action. In order to find the 
correction to the tachyon potential one should consider the partition 
function on the annulus for a single D9-branes which can be obtained under
certain condition which has to be imposed for $x$ and $y$. This is what we are
 going
to compute in the next section. 

The spectrum of those open strings with different boundary conditions can also 
be 
obtained by setting, for example, $x=0$. In this case the partition function
(\ref{TT}) reads
\be
Z[y]=Z_{\rm disk}[y]\int_0^1 {da \over a}
\frac{\prod_{k=1}^{\infty} (1+C_k)}{\prod_{r={1\over 2}}^{\infty} (1+C_r)}\, .
\label{DIFB}
\ee
Using the expansion of (\ref{DIFB}) 
around $y=0$, one can find the corresponding string 
action up to two-derivative 
\be
S=T_9\int d^{10}X \int_{0}^{1} {da\over a} 
\prod_{k=1}^{\infty} \frac{1-a^{2 k-1}}{1-a^{2 k}} 
e^{-{1 \over 4}T^2}\left[1+2 \ln 2\; 
\partial^{\mu}T\partial_{\mu} T\right].
\ee
Finally we have the following expression for the partition function before
and after tachyon condensation (for the bosonic string
see \cite{TSU}) 
\bea
Z&\sim & \int_{0}^{1} {da\over a}\prod_{k=1}^{\infty}\frac{1-a^{2k-1}}
{1-a^{2k}},\;\;\;\;\;\;\;\;\;\;\;\;\;\;\;\;\; {\rm NN}\; (x=y\rightarrow 0)\; ,
\cr &&\cr &&\cr
Z&\sim & \int_{0}^{1} {da\over a}(\ln a)^{-{1\over 2}}
\prod_{k=1}^{\infty}\frac{1-a^{2k-1}}
{1-a^{2k}},\;\;\;\;\; {\rm DD}\; (x=y\rightarrow \infty)\; ,
\cr &&\cr &&\cr
Z&\sim & \int_{0}^{1} {da\over a}\prod_{k=1}^{\infty}\frac{1+a^{2k-1}}
{1+a^{2k}},\;\;\;\;\;\;\;\;\;\;\;\;\;\;\;\;\; {\rm ND}\; (x=0, y\rightarrow 
\infty)\; .
\eea

\section{One-loop Tachyon Action}

So far we have used the annulus partition function in order to study the
spectrum of the
open string stretched between two parallel D-branes. As we have
already mentioned, having an annulus would naturally lead to a system with two
branes. Nevertheless, under a certain circumstances one can also 
have a system with
a single D9-branes. 
Regarding the fact that under exchanging of $y\leftrightarrow ax$ we go from one 
boundary to the other one (see section 2), setting $y=ax$
would mean that we are identifying two boundaries of 
the annulus and this, somehow, means
that both boundaries represent a single D9-brane. In this case, 
we have $C^{-1}_k=a^{-2k}\left(\frac{k+y}{k-y}\right)^2-1$. Therefore, one finds
\be
\frac{\partial {\rm ln}Z[y, a]}{\partial y}=2
{\partial\over\partial y}\ln
\left[|y^2\ln a-2y |^{-{1\over 4}} 
\prod_{k=1}^{\infty}\frac{[(k+2 y)^2-(k-2 y)^2
a^k]^{1/2}}
{(k+y)^2-(k-y)^2a^{2k}}\right]\; ,
\ee
where the factor of two is because of the contribution of two boundaries.
Doing the same computation as that in the previous section, we get
\be
Z[y]=Z_0\int_0^1 {da\over a}\;\frac{F^2(y)}{\sqrt{|y^2\ln a-2y|}}
\;\frac{\prod_{k=1}^{\infty} (1+C_k)}{\prod_{r={1\over 2}}^{\infty} (1+C_r)}\, .
\ee
By making use of
 this expression for the partition function on the annulus, the 
one-loop correction to the string field
theory action for the tachyon can be obtained as follow
\be
S=T_9\int d^{10}X e^{-{1\over 2}T^2} \int_0^1 {da \over a}
\frac{F^2\left(\partial_{\mu}T\partial^{\mu}T\right)}{\sqrt{|1-{\ln a\over 2}
\partial_{\mu}T\partial^{\mu}T|}}\;
\frac{\prod_{k=1}^{\infty} [1+C_k(\partial_{\mu}T\partial^{\mu}T)]
}{\prod_{r={1\over 2}}^{\infty} [1+C_r(\partial_{\mu}T\partial^{\mu}T)]}\, ,
\label{ONEL}
\ee
where
\be
C_n(\partial_{\mu}T\partial^{\mu}T)=\frac{a^{2n}}{\left(\frac{n+
\partial_{\mu}T\partial^{\mu}T}{n-\partial_{\mu}T\partial^{\mu}T
}\right)^2-
a^{2n}}\;\;\;\;\;\;\;\;\;\;\;\; n=r,\; k\; .
\ee 
In particular the one-loop correction to the tachyon potential is
\be
V_{\rm one-loop}(T)=T_9\; e^{-{1\over 2}T^2} \int_0^1 {da \over a}
\prod_{k=1}^{\infty}\frac{1-a^{2k-1}}{1-a^{2k}}\; .
\ee
As we see, the annulus correction to the potential does not change the shape 
of the potential given by the disk computation \cite{KMM2}, i.e. it has
extremum at $T=0,\infty$. Nevertheless the unstable D9-brane tension has to be
renormalized as the above integral divergences at $a=0$. More precisely, setting
$\lambda=g_se^{-T^2/4}$ \cite{KMM} and adding this 
result to that of disk computation \cite{KMM2}, one finds
\be
V(T)=T_9 e^{-{1\over 2}T^2}\left(\frac{1}{\lambda}+
\int_0^1 {da \over a}
\prod_{k=1}^{\infty}\frac{1-a^{2k-1}}{1-a^{2k}} \right)\; .
\ee
One can now absorb the infinite part of the modulus integral by renormalizing
the effective string coupling 
\be
V(T)=T_9 e^{-{1\over 2}T^2}\left(\frac{1}{{\bar \lambda}}+\Lambda \right)\; .
\ee
where $\bar \lambda$ is the renormalized effective coupling and 
$\Lambda$ is the finite part of the modulus integral.

\section{Conclusion}

In this letter we considered the superstring field theory on the annulus and 
computed the partition function of the world-sheet action on the annulus
with the linear tacyhon perturbation. By making use of the partition 
function, we have been able to write down the one-loop correction to the
string field action of the tachyon. In this computation, we assumed that 
$Z[x,y]=S[x,y]$, as we have
in the string field theory on the disk. But it remains to be argued, if this
assumption is correct beyond the disk approximation. Looking at the explicit 
derivation of the relation between string action and the partition function in 
the superstring case \cite{M, NP}, it is not a priori obvious that this should
be the case, as their results depend on the explicit form of the correlation 
function of matter and ghost fields on the disk. Nevertheless, since 
the partition function we found is monotonically decreasing,
hopefully setting $S=Z$ comes true after all !

Using the one-loop computation, it seems that we would get the following 
perturbative expansion for the partition function
\be
Z=e^{-{1\over 2}T^2}\left({1\over \lambda} Z_1+Z_0+\cdots\right)
\label{EXP}
\ee
where $Z_{\chi}$ is obtained from the path integral on surface of 
Euler number $\chi$ and no handles. This form has recently been suggested
in \cite{KMM}. We would like to note that, in the early version of this paper, 
the one-loop
correction to the tachyon potential had been extracted from equation 
(\ref{HAD}) 
and therefore we had concluded that the genus expansion is not governed 
by the effective string coupling defined in terms of the tachyon (\ref{EXP}), 
as has been discussed in \cite{KLS}. In fact, as we mentioned in the previous 
section, the correction to the tachyon action has to be read from (\ref{ONEL})
rather than (\ref{HAD}).

\vspace*{1cm}
{\bf Note added}: While typing the paper we received \cite{YY2} where the
same subject has been studied, though, with a minor different result. This
is because the authors of \cite{YY2} have missed the modulus dependence
of the integration constant. Moreover it seems that the $\zeta$-function 
regularization has not correctly applied to find the partition function. In
particular for the case of $y_a=y_b\equiv y$ in their notation, the
partition function is not monotonically decreasing (see equation (54) in
\cite{YY2}) and in fact it grows like $4^{y}$ as $y\rightarrow \infty$. 
Of course, this is not what we would
like to have as we are going to identify the partition function with the
boundary action.

\vspace*{1cm}

{\bf Acknowledgments}

I would like to thank J. de Boer for collaboration at the early stage of 
this work and for useful discussions and comments. I am grateful to  Y. Yang 
for discussions and his comment on 
the early draft of the paper. I would also like to thank R. Dijkgraaf, 
D. Kutasov, T. Suyama and C. Vafa for useful discussions and comments.

\end{document}